# Enhancing semantic expressivity in the cultural heritage domain: exposing the Zeri Photo Archive as Linked Open Data

Marilena Daquino, marilena.daquino2@unibo.it
Department of Classical Philology and Italian Studies
University of Bologna, Bologna, Italy

Francesca Mambelli, francesca.mambelli@unibo.it
Federico Zeri Foundation
University of Bologna, Bologna, Italy

Silvio Peroni, silvio.peroni@unibo.it
Digital And Semantic Publishing Laboratory
Department of Computer Science and Engineering
University of Bologna, Bologna, Italy

Francesca Tomasi, francesca.tomasi@unibo.it
Department of Classical Philology and Italian Studies
University of Bologna, Bologna, Italy

Fabio Vitali, fabio.vitali@unibo.it
Digital And Semantic Publishing Laboratory
Department of Computer Science and Engineering
University of Bologna, Bologna, Italy

## ABSTRACT

Describing cultural heritage objects from the perspective of Linked Open Data (LOD) is not a trivial task. The process often requires not only choosing pertinent ontologies, but also developing new models that preserve the most information and express the semantic power of cultural heritage data. Indeed, data managed in archives, libraries and museums are complex objects themselves, which require a deep reflection on even non-conventional conceptual models. Starting from these considerations, this paper describes a research project: to expose the vastness of one of the most important collections of European cultural heritage, the Zeri Photo Archive, as Linked Open Data. We describe here the steps we undertook to this end: firstly, we developed two *ad hoc* ontologies for describing all the issues not completely covered by existent models (the *F Entry* and the *OA Entry Ontology*); then we mapped into RDF the descriptive elements used in the current Zeri Photo Archive catalog, converting into CIDOC-CRM and into the two new aforementioned models the source data based on the Italian content standards Scheda F (Photography Entry, in English) and Scheda OA (Work of Art Entry, in English); and finally, we created an RDF dataset of the output of the mapping that could show a result capable of demonstrating the complexity of our scenario.

**RASH:** https://w3id.org/people/essepuntato/papers/zeri-and-lode-jocch.html

# INTRODUCTION

Libraries, archives and museums are facing a substantial transformation in the management of cultural heritage data. In particular, the attempt to express the latent semantic power of data – by creating relations and interconnections between different kinds of cultural objects owned by similar cultural institutions – is considered a shared approach in the community. The idea to define an infrastructure for sharing metadata about such objects is needed in order to create comprehensive tools for researchers, to provide innovative environments for users' learning, and, finally, to increase the impact of culture on society. Cultural heritage institutions are dealing with two urgent issues: on the one hand, they need to provide a complete and exhaustive semantic description of their data; on the other hand, they have to open up their data to interchange, interconnection and enrichment.

The urgent need for such scenarios has been addressed in the context of a particular project, *PHAROS: An International Consortium of Photo Archives* [24]. PHAROS is one of the first concrete steps towards the creation of an actual digital infrastructure of the notable photographic archives of works of art in Europe and the United States of America. The Consortium enables active collaboration between the institutions responsible for fourteen photo archives so as to create a common platform for research on images and metadata of Western and non-Western works of art in all media. One of the most important institutions included in the Consortium is the Zeri Foundation (http://www.fondazionezeri.unibo.it/en). Its Photo Archive is planned to be one of the first assets to be included in PHAROS.

In this article we introduce a project called *Zeri and LODE*, which addresses the enhancement of existing metadata collections of the Zeri Photo Archive by means of Semantic Web technologies, in order to make them easily discoverable, interchangeable, and interlinked with other existing and relevant sources.

Our sources were the collections owned by one of the most important art historians of the twentieth century, Federico Zeri (1921-1998). His extensive library of art books, auction catalogues and individual photos of monuments and artworks were the main tools of his work. He began to collect them in the 1940s and, over time, made them into "the world's largest private archive of Italian paintings", which became an essential reference work for the historical sequencing of out-of-context works. At the time of his death, the archive included more than 46.000 volumes, 37.000 auction catalogues, 60 periodicals and 290.000 individual photographs. In order to preserve his bequest and put it to best use, the Zeri Foundation was set up in his name at the University of Bologna, and it has come to be recognized as one of the most important research and training centres for art historians in the world.

Among its activities, the Zeri Foundation undertook the cataloguing of Zeri's repository, employing to that end two Italian metadata content standards, *Scheda F*, for *Scheda di fotografia* (*photograph*, available at http://www.iccd.beniculturali.it/index.php?it/473/standard-catalografici/Standard/10) and *Scheda OA*, for *Scheda Opera d'Arte* (*work of art*, available at http://www.iccd.beniculturali.it/index.php?it/473/standard-catalografici/Standard/29). We will use the original names Scheda F and Scheda OA below to refer to aforementioned content standards, while the English translation *F Entry* and *OA Entry* will be used to refer to metadata documents recorded by adopting the aforementioned content standards, and the resulting ontologies, named F Entry and OA Entry as well. Both content standards are issued by the *ICCD* (*Istituto Centrale per il Catalogo e la Documentazione*), Central Institute for the Cataloguing and Documentation), which is part of the Italian Ministry of Cultural Heritage. The ICCD coordinates research into the definition of cataloging standards for all kinds of cultural objects, including all those belonging to the archaeological, architectural, artistic and ethno-anthropological areas of interest to the ministry.

*Scheda F* includes more than 300 fields recording information about photographers, the history of photo production, publication, preservation, changes of location, exhibitions, materials and techniques, reference bibliographies, attached documentation and technical data derived from the philological analysis of the item. In the context of the Zeri Photo Archive, the *Scheda OA* refers to the content of photography, i.e. the work of art. In particular, it includes more than 200 fields that offer a detailed description of the object depicted and provides similar information to that described in *Scheda F*.

The work of cataloguing Zeri's collection in compliance with these two Italian standards has resulted in the Zeri Photo Archive catalogue, which is stored in a relational database and is accessible through a web interface (http://catalogo.fondazionezeri.unibo.it).

Because of the national and international relevance of this photo archive, the complexity recorded in its heterogeneous catalogue, and the explicit request from the PHAROS Consortium, we have started to work on these data as a use case for defining a common model between institutions that could use different descriptive standards for photographs. In particular, our main goal was the development of appropriate Semantic Web models and technologies so as to provide a representation of the Zeri Photo Archive as Linked Open Data (LOD).

The PHAROS Consortium suggested the CIDOC-CRM conceptual reference model [17] as the required representational framework for data to be shared within the consortium, given its explicit characterization for cultural heritage entities and its widely recognized international standing. CIDOC-CRM has been widely adopted in Italy in the recent years and was considered by PHAROS members to be a good starting point for addressing dialogue between data sources – even if no photo archive had been described by means of it yet. Thus, the original plan was to map the Scheda F and Scheda OA fields with CIDOC-CRM properties, so as to represent the metadata of the Zeri Photo Archive into a machine-readable form according to the required standard.

Unfortunately the full mapping was not possible, since many properties of photos and works of art described by Scheda F and Scheda OA, that are necessary for the full and complete description of Zeri's catalogue, do not have a natural representation in CIDOC-CRM. For instance, a clear representation of subjective attributions (e.g. authorship attributions, the object's appellation, dating, and the influence between works) should be handled by means of dedicated entities, and should also consider the provision of data integration among stakeholders – who may otherwise catalogue the same object providing contradictory information. Moreover, there should be a precise representation of the network of people and organizations involved in the lifecycle of such cultural objects with different roles, together with an explicit representation of the influence a cultural object may have on the conception of another one (e.g. a copy or a derivative work). While all these data are crucial for enabling further relations between otherwise undiscoverable information, they cannot be represented in CIDOC-CRM without ambiguities.

So as to address all the aforementioned issues, we focussed on the creation of:

1. two ontologies, i.e. the F Entry Ontology and the OA Entry Ontology, to represent all missing information from CIDOC-CRM;

2. two documents, i.e. an F entry and an OA entry document, exemplifying the mappings and alignments from Scheda F and Scheda OA to CIDOC-CRM and our ontologies, so as to provide a complete representation of the description requirements for photographs and works of art;

3. an RDF dataset, published as Linked Open Data, describing a large number of the

available F/OA entries in Zeri's catalogue according to the aforementioned ontologies.

Even if the new entities introduced in the ontologies we have developed mainly come from terms of two specific national standards (i.e. Scheda F and Scheda OA), entities and properties described in the ontologies are of widespread application and are, to our knowledge, the largest and most complete model for describing photographs, works of art, and related documentation. Other archives planning to describe collections of photos in CIDOC-CRM could find our experience in working with Zeri's catalog very useful for their own representation purposes.

This paper extends and details works already introduced in [13]. In addition to what is described in that paper, we provide a thorough revision of the F Entry Ontology triggered by the OA Entry Ontology development. This includes the addition of topics that had not been tackled previously: a full description of works of art depicted in catalogued photographs, the development of aspects that were deemed secondary when first developing the F Entry in isolation, and the uniformity of choices for RDF representation of both work types (photos and works of art). Then we proceeded with the mapping into RDF of terms belonging to Scheda F and Scheda OA used in the Zeri Photo Archive, according to aforementioned ontologies. Finally, we realized the Linked Open Data publication of the first subset of the Zeri catalogue, including photos of artworks from the fifteenth and sixteenth centuries. All stages of the process, from the mapping to the ontologies development and the LOD publication, have been shared with the ICCD and are still under discussion in order to define a unique general model that could be proposed to all the Italian institutions that have to manage this kind of data.

The rest of this paper is organised as follows. In RELATED WORKS we introduce some related works on these topics. In METHODOLOGY AND ONTOLOGY DEVELOPMENT we present the methodology we adopted for the development of the ontologies, which are introduced in THE F ENTRY ONTOLOGY AND THE OA ENTRY ONTOLOGY. In MAPPING F/OA ENTRY TO RDF we discuss the alignments among the two content standards (i.e., Scheda F and Scheda OA) into RDF. In THE ZERI PHOTO ARCHIVE RDF DATASET we describe the LOD dataset we have created that includes all the RDF data obtained from converting the Zeri Photo Catalogue. Finally, in CONCLUSIONS we sketch out some future works.

# RELATED WORKS

As said in the introduction, the digital cultural heritage scenario is facing a strong transformation. In particular, publishing strategies of cultural heritage data are changing the traditional methodologies. Cleaning, reconciliation, enrichment and linking [29] are the new keywords in the domain of libraries, archives and museums. The Semantic Web, and the Linked Open Data theory and practice, have determined a revolution in data production and access, and both machines and final users are experimenting with new ways of acquiring knowledge from URIs and typed RDF links (http://linkeddata.org).

Many projects, e.g. LODLAM (http://lodlam.net) and OpenGlam (http://openglam.org), demonstrate how much the community is growing around these themes (for an updated survey see the Task Force activity on LOD: https://www.w3.org/wiki/SweoIG/TaskForces/CommunityProjects/LinkingOpenData).

Datasets are increasing in number (see the datahub.io classification) and the LOD cloud grows larger every day. Together with the publication of LOD collections, new vocabularies are being modelled (see LOV, Linked Open Vocabularies, http://lov.okfn.org/dataset/lov/), in order to address the needs of the cultural heritage domain. The scientific literature about methods of producing LOD shows the direction of research around these themes, towards authorities reconciliation and linking, but also ontologies reuse and development, which are the basis for realizing the idea of LOD.

The need to convert flat traditional element schemes, such as the above-mentioned Scheda F and Scheda OA, into ontology terms forced us to develop new models as a result of a reflection on existing conceptualizations and their shortcomings. The adoption of models that were widely appreciated in the specific domain of cultural heritage is the first obvious step, but we also considered other models already capable of addressing additional important issues.

As described in the next sections, many ontologies represent important facets in our ontology development process. Each chosen ontology was used to cover one aspect of our heterogeneous scenario and were therefore first reused, but then also extended. The SPAR ontologies (http://www.sparontologies.net), mostly used in academic literature, were necessary in order to describe the documentation used in the cataloguing process. But, even when describing the catalogue entries using SPAR, we aimed to also make them available as bibliographic sources. In particular, FaBiO [22] was chosen for managing bibliographies according to the FRBR approach [14], CiTO [22] was reused for describing the different citations that cataloguers provide to support their attributions, and PRO [23] was fundamental for documenting people's role in photographic, arts, publishing and cataloguing domains. HiCO [6], was created by extending PROV-O [16], in order to describe the interpretation process relative to subjective attributions, a fundamental aspect of our conceptualization. And of course CIDOC-CRM [17] was finally used in order to describe all the features directly related to catalogued objects, so as to guarantee semantic interoperability between stakeholders.

The need for ontological interconnection represents a trend: as Europeana demonstrates, in particular within the EDM (Europeana Data Model) [9], the variety of existing vocabularies in the cultural heritage domain creates the need to reuse and align models [8]. The mapping to CIDOC-CRM forced us to analyse methodologies in merging and mapping, as technical processes, theoretical activities and common, shared, approaches in the cultural heritage domain [11].

In addition to the ontologies, we made use of authorities, controlled vocabularies and other datasets (e.g. VIAF, DBpedia, Getty AAT and ULAN, Wikidata, Geonames), in order to define appropriate links for both mutual connections and alignment (a list in [15]).

Other existing datasets in the artworks domain were finally analysed (e.g. Smithsonian Art Museum, http://americanart.si.edu/collections/search/lod/about/; Yale Center for British Art, http://britishart.yale.edu/collections/using-collections/technology/linked-open-data).

## METHODOLOGY AND ONTOLOGY DEVELOPMENT

Modelling complex information provided by Scheda F/OA content standards meant managing manage different knowledge domains, i.e. bibliographic, archival, photographic, arts and cultural heritage domains. Thus, our priority was to enable the coexistence of already exixtent and available heterogeneous models, developed for describing one of the aforementioned domains, as big modular consistent ontologies for describing all of the entities involved in Zeri's Catalogue.

Usually, cultural institutions adopt the most suitable vocabulary for their purpose, without paying attention to  the way similar concepts are modelled in other relevant domains, delaying the possibility of aligning them as possible future work. The same object may be described in different ways according to the data creator's cultural background. For example, Arts and Library Sciences conceive and apply differently the Functional Requirements for Bibliographic Records (FRBR) conceptual model [1] [2] [5] [26] into their object description. Cataloguing artworks generally entails aggregating under the same label (i.e. type of work) the functional, formal, and morphological features of the object, by considering these issues inseparable from the conception of the work itself, i.e. the FRBR Work. This is a peculiarity of the arts domain. In the bibliographic domain, however, such definition may be formalized by considering these features at levels other than those of the Work itself, e.g. levels of Expression and Manifestation. Thus the RDF representation of a painting may significantly vary when defined in one of the two aforementioned domains, which will affect the way that data can be integrated.

In addition to purely theoretical and design related considerations, the way different data providers label their data should be taken into account as well. For instance, cataloguers use specific and often customised criteria for recording authorship attributions, e.g. by means of their own classification system, that are not immediately sharable by other stakeholders without a reconciliation into a controlled vocabulary. Thus, one of the main aspects we had to address was providing flexible mechanisms [20] to record the particular criteria adopted by a cataloguer for cataloguing a work, still allowing their evolution over time if new cataloguing guidelines were to be proposed and to substitute for to the old ones, without any loss of precious information. In addition, keeping track of the multiple and even inconsistent interpretations in cataloguing was something that had to be handled properly as well.

In order to provide a model to describe the aforementioned cases, we also wanted to rely on existing standards and models, instead of reinventing everything from scratch. The use of FRBR-compliant ontologies (i.e., SPAR Ontologies) for modelling the bibliographic and documental aspects of our domain, the HiCo Ontology to provide attributions of provenance information, and CIDOC-CRM, i.e. the international standard for describing cultural artefacts, appeared to be quite useful for providing a first, even if still partial, description of works of art and photographs and their related data.

We initially tried to work out if an alignment of Scheda F/OA elements to entities the CIDOC-CRM model was possible. Usually, the alignment to existing models is not a straightforward operation and could require several iterations and integrations to be prepared properly. We conducted a brief analysis of the CIDOC-CRM data model, which revealed that several concepts we needed in order to properly describe our domain were not available in such a reference model.

Therefore, our development first defined terms and relations so as to accomplish the full description of our domain, and then we refactored and aligned them into CIDOC-CRM. We also developed two OWL 2 DL ontologies [19] for expressing concepts and relations of the main descriptive areas of the Italian content standards Scheda F/OA in RDF that were lacking in CIDOC-CRM. These were released with Creative Commons licenses in order to foster broad reusability. The approach we followed was mainly data-centric. In fact, these ontologies had to properly describe all the data represented in a real and representative dataset compliant with F/OA standards, i.e. the Zeri Photo Archive catalogue. Thus, the use of this dataset allowed us to develop ontologies whose entities had meaningful names, and to provide real examples of their usage at every stage of the development process.

The methodology we adopted was the Simplified Agile Methodology for Ontology Development (*SAMOD*) [21]. SAMOD is an agile methodology, developed by starting from guidelines proposed in well-known and existing ontology development methodologies such as [12] [28], which allowed us to develop the model by means of several small and iterative steps, and to create documentation by using examples of data. In particular, this methodology required us to consider small issues iteratively, and provided us with a way to test our under-development model immediately on a real use case in any phase of the development. The logical consistency of the model is ensured by the use of a reasoner on it, while data consistency is achieved by testing the model on test cases. Moreover, the coherency of the model with the domain is assessed by the domain expert, who checks the correctness of the vocabulary, the adopted thesauri, and the relations.

Good practices in ontology development were respected to ensure semantic interoperability and to facilitate the model's reuse: defined concepts and relations were refactored, directly reusing the most appropriate external entities from established and well known ontologies, or by aligning them to those of other models.

Documentation of both the F Entry and OA Entry ontology development processes is available online, at http://www.essepuntato.it/2014/03/fentry/samod and http://oaentry-ontology.sourceforge.net/samod/OAdevelopment.zip respectively.

In the next section we provide a complete description of covered aspects and their formalization as ontological entities, accompanied by exemplars of their usage.

# THE F ENTRY ONTOLOGY AND THE OA ENTRY ONTOLOGY

As written above, the main objective of our project is to open the Federico Zeri Foundation data to the world according to the standard CIDOC-CRM model, which was chosen as the *lingua franca* for sharing data between stakeholders. However, CIDOC-CRM lacks some particular entities belonging to the photography and arts domains, as well as several relations that in our opinion would make a true integration possible between museums, as well as between museums and other domains.

Therefore, we extended the CIDOC-CRM model by creating two complementary ontologies, the F Entry Ontology and the OA Entry Ontology, that are described in the following subsections and that allow the modelling scenarios to be addressed which are not properly addressed by CIDOC-CRM. The actual mapping between CIDOC-CRM and the F/OA content standards is described in MAPPING F/OA ENTRY TO RDF.

The first ontology developed by using the above described methodology is the F Entry Ontology, which deals with issues related to the photography domain and several aspects that are in common with the OA Entry content standard for describing works of art. The latter ontology, the OA Entry Ontology, is thus very similar to the former one, and basically provides additional entities and relations proper to a comprehensive representation of the arts domain. The development of the OA Entry Ontology also resulted in a revision of the F Entry Ontology, so as to provide a specular description of same issues when appropriate. For the sake of brevity, common aspects of the two ontologies are detailed only once, in The F Entry Ontology (FEO).

## *The F Entry Ontology (FEO)*

The current version of F Entry Ontology (FEO) revises the previous version introduced in [13], and it is available online at http://www.essepuntato.it/2014/03/fentry. The Graffoo diagram [10] of FEO in Illustration 1 provides an overview of its classes – i.e. the yellow rectangles –, object properties – i.e. the blue dotted lines beginning with a solid circle and ending with a solid arrow –, and assertions among classes – i.e., the black lines ending with a solid arrow (the full specification of Graffoo graphical elements is available at http://www.essepuntato.it/graffoo/specification/current.html).

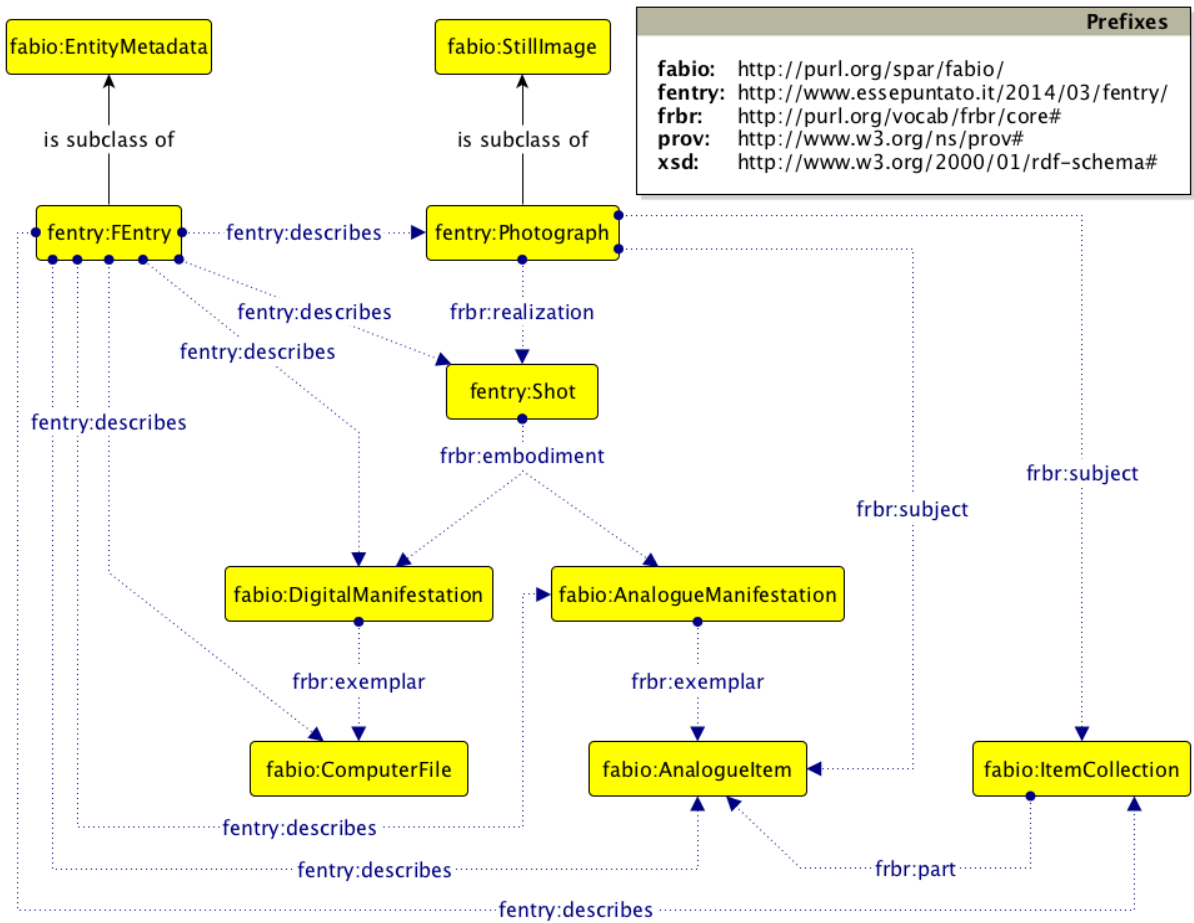

*Illustration 1: The Graffoo diagram of the F Entry Ontology main classes and properties.*

FEO introduces classes and properties needed to characterize three main concepts: the photograph, the subject portrayed in the photograph and the F Entry describing the photograph and its subjects. Each of the aforementioned entities is characterised in terms of FRBR. In particular:

- the photograph is represented as an FRBR Work when describing its essence, as an FRBR Expression when dealing with information about its realization (i.e. the shot), as an FRBR Manifestation when describing each tangible form of the photograph, and as FRBR Item for each individual copy with different features;

- in the Zeri Photo Archive the portrayed subject of a photograph is always a concrete work of art. Therefore, it can be represented as an FRBR Work when defining its essence, and as FRBR Item when describing the concrete object. The entities representing this sort of subject were refined when the OA Entry Ontology was created, which is explained in detail in The OA Entry Ontology. When the subject is not identified as a concrete object, a place or an agent, it can be simply considered as a term from an open/controlled vocabulary or thesaurus.

- the F Entry is a work containing metadata about a photograph and its cataloguing; it is subject to several revisions, each of which is related to responsible entities (i.e. cataloguers and supervisors); we are not interested in how it is preserved, what formats it is in and how many copies there are. Thus an F Entry is represented as an FRBR Work when describing its creation and as an FRBR Expression when describing its contents and revisions.

Existing ontologies have been imported into FEO so as to provide a precise description of specific aspects of the domain in consideration. In particular, we imported the FRBR-aligned Bibliographic Ontology (FaBiO, prefix *fabio*) [22], the Publishing Roles Ontology (PRO, prefix *pro*) [23], the Historical Context Ontology (HiCO, prefix *hico*) [6], and the Citation Typing Ontology (CiTO, prefix *cito*) [22]. In addition to terms from these ontologies, terms from an OWL 2 DL version of FRBR (prefix *frbr*, http://purl.org/spar/frbr, imported by FaBiO) are also used, so as to represent hierarchical and associative relations between the main entities, as well as terms defined in the Provenance Ontology (PROV-O, prefix *prov*) [16].

In the next sections we describe in more detail how we used and extended the imported ontologies – i.e., FaBiO, PRO, CiTO, HiCO – to describe the different aspects of the domain. The complete RDF example of the excerpts provided in the following sub-sections is available at http://dx.doi.org/10.6084/m9.figshare.3175252.

## Extending FaBiO to define the cultural object

FaBiO was originally developed for describing bibliographic entities according to the FRBR conceptual model. It mainly addresses issues related to published texts, by introducing wide taxonomies of possible kinds of works, expressions, manifestation and items, also defining them by means of several object properties domain and range constrains. It was therefore a good place from which to start, which we refined in order to model our main entities (i.e. F/OA entries, photographs, and works of art), so as to enable a deeper characterization of them according to well-known models already in use.

An F Entry can describe a photograph in each phase of its lifecycle: from its creation (i.e. the essence of the photograph) to its realization (i.e. the shot), from its development into a visible image (i.e. negative, positive, slide, digital image) to its publishing and reproduction. Each phase of the lifecycle of the photograph corresponds to an instance of a class defined in terms of FRBR as defined in FaBiO.

An F Entry is a document containing metadata about a photograph and its cataloguing. Therefore it is defined in terms of FRBR Work (i.e. an instance of the class *fentry:FEntry*, subclass of *fabio:EntityMetadata*). Subjects of the entry are considered as FRBR Works as well, so that includes the photograph (i.e. an instance of the class *fentry:Photograph*, subclass of *fabio:StillImage*) and, in the Zeri Photo Archive context, the work of art portrayed in the photograph (i.e. an instance of the class *fabio:ArtisticWork*).

The object property *fentry:describes* links an instance of the class *fentry:FEntry* to instances of the proper class defining its subject, among which there are classes *fentry:Photograph* and *fabio:ArtisticWork*. For example, the natural language scenario "an F Entry describes the photograph portraying the painting called *Jesus's baptism*", may be expressed, in Turtle syntax, as follows:

```
:fentry-72486 a fentry:FEntry ;
      fentry:describes
            :jesus-baptism-photo-work , :jesus-baptism-photo-item ,
            :jesus-baptism-work , :jesus-baptism-item .
:jesus-baptism-photo-work a fentry:Photograph  .
:jesus-baptism-photo-item a fabio:AnalogItem .
:jesus-baptism-work a fabio:ArtisticWork .
:jesus-baptism-item a fabio:AnalogItem .
```

Thus, the shot (i.e. an instance of the class *fentry:Shot*, subclass of *fabio:Expression*) is a realization of a photograph, which can take several forms when developed, principally distinguished by means of their analogue/digital formats (i.e. instances of the classes *fabio:AnalogManifestation* or *fabio:DigitalManifestation*) and can be individually described (i.e. an instance of the classes *fabio:AnalogItem* or *fabio:DigitalItem*).

For example, the natural language scenario "The shot of the photograph portraying the *Jesus's baptism* painting that had been taken by Brogi before 1940 was published by himself in 1940", can be expressed, in Turtle syntax, as follows:

```
:jesus-baptism-photo-work
      frbr:realization :jesus-baptism-photo-shot .
:jesus-baptism-photo-shot  a fentry:Shot ;
      frbr:embodiment :jesus-baptism-photo-positive .
:jesus-baptism-photo-positive a fabio:AnalogManifestation .
```

## Using PRO to describe the lifecycle of the object

PRO allows one to describe scenarios in which agents hold roles with respect to a particular time and context, e.g. the fact that a person was (property *pro:holdsRoleInTime*) the owner (the role in consideration, specified through the property *pro:withRole*) of a certain photograph (the context, linked with the property *pro:relatesTo*) from the 27th of November 2002 to the 16th of June 2015 (the time interval, specified by means of the property *tvc:atTime*). An instance of the class *pro:RoleInTime* is created every time we need to specify these kinds of roles.

For instance, in our context, the role of Brogi as photographer is held on the shot, i.e. an FRBR Expression, and the publisher role is held on the positive of the photograph, i.e. an FRBR Manifestation. These relations can be represented as follows:

```
:brogi a foaf:Agent ;
      pro:holdsRoleInTime :brogi-photographer-jesus-baptism-photo-shot ;
      pro:holdsRoleInTime :brogi-publisher-jesus-baptism-photo-positive .
:brogi-photographer-jesus-baptism-photo-work a pro:RoleInTime ;
      pro:withRole scoro:photographer ;
      pro:relatesTo :jesus-baptism-photo-shot ;
      tvc:atTime :jesus-baptism-photo-shot-date .
:brogi-publisher-jesus-baptism-photo-positive a pro:RoleInTime ;
      pro:withRole pro:publisher ;
      pro:relatesTo :jesus-baptism-photo-positive ;
      tvc:atTime :jesus-baptism-photo-publishing-date .
```

There may be some situations in which the creator and the realizer of the shot are not the same person. Therefore the main photographer is described by means of the provided CIDOC-CRM terms for representing the creation of a work (explained in next MAPPING F/OA ENTRY TO RDF) and the terms belonging to PRO (or to ScoRO, http://purl.org/spar/scoro, which extends PRO with additional roles) for describing roles other than that of the *creator*.

## Using HiCO to provide the provenance of assertions

HiCO has been developed to describe the interpretative process underlying questionable information, by means of a provenance statement characterizing the entity that may be contradicted by another data provider. Each piece of questionable information is thus defined by an individual of the class *hico:InterpretationAct*, which allows one to specify the scope (property *hico:hasInterpretationType*) for the current interpretation, and the criteria (*hico:hasInterpretationCriterion*) underlying such interpretative choice.

In addition, each instance of *hico:InterpretationAct* is also linked (property *hico:isExtractedFrom*) to the text source where such questionable information is stated in natural language. In our domain, such source is always the content of the F Entry, subclass of *fabio:MetadataDocument* – i.e., an FRBR Expression. Thus, it is worth mentioning that the creator (property *prov:wasAssociatedWith*, defined in PROV-O [16]) of the RDF statements describing an interpretation act and the author of the original text that they were extracted from are not necessarily the same person. Finally, the relation between the agent holding a certain role (an individual of the class *pro:RoleInTime*, described in the previous sub-section) and the RDF-defined interpretation act is introduced by means of the object property *prov:wasGeneratedBy*.

For instance, consider the following natural language text derived from an existing F Entry and describing some information about the photograph portraying the painting *Jesus's baptism*, by Leonardo da Vinci:

> the attribution of Brogi as the publisher of the photograph portraying the *Jesus's baptism* painting, was motivated by a formal analysis of the photograph itself, which revealed on its verso an inscription naming Brogi as publisher.

This natural language scenario can be expressed in RDF by using the ontological entities introduced above, as follows (in Turtle syntax):

```
:brogi-publisher-jesus-baptism-photo-positive
       prov:wasGeneratedBy :jesus-baptism-photo-publisher-attribution .
:jesus-baptism-photo-publisher-attribution a hico:InterpretationAct ;
       hico:hasInterpretationType :role-attribution ;
       hico:hasInterpretationType :zeri-preferred-attribution ;
       hico:hasInterpretationCriterion :formal-analysis ;
       hico:hasInterpretationCriterion :inscription ;
       hico:isExtractedFrom :fentry-72486-expression ;
       prov:wasAssociatedWith :crr-mm .
:role-attribution a hico:InterpretationType .
:zeri-preferred-attribution a hico:InterpretationType .
:formal-analysis a hico:InterpretationCriterion .
:inscription a hico:InterpretationCriterion .
:fentry-72486-expression a fabio:MetadataDocument .
:crr-mm a foaf:Agent .
```

In the excerpt, the instance `:zeri-preferred-attribution` is provided in order to distinguish the current interpretation as the one chosen by the cataloguing institution from other possible alternative interpretations specified.

## Using CiTO for relating documents and attributions

The relation between an interpretation and a heterogeneous source supporting a certain cataloguer's interpretation can be defined as proper (even implicit) citation. CiTO allows one to mark citation links between authors, and to specify the intent of such citation by providing a wide set of object properties relating citing and cited entities. To this end, we can use the object properties provided in CiTO for linking an individual of the class *hico:InterpretationAct* to the original textual interpretation from which the interpretation act was derived.

For instance, the fact that the cataloguer is citing as evidence an inscription recognized on the verso of the photograph, can be represented by means of the object property *cito:citesAsEvidence* as follows:

```
:jesus-baptism-photo-publisher-attribution
       cito:citesAsEvidence :jesus-baptism-photo-verso .
```

## *The OA Entry Ontology*

The Scheda F content standard provides just a few elements regarding the work of art that may be portrayed in a photograph. However, there exists another content standard, i.e. the Scheda OA, that aims to be an exhaustive reference document providing a complete description of any work of art – not only those portrayed in photographs. For this reason, all of the aspects peculiar to the work of art portrayed in a photograph have been modelled in the OA Entry

Ontology (available at http://purl.org/emmedi/oaentry). The Graffoo diagram in Illustration 2 provides an overview of its main classes and properties.

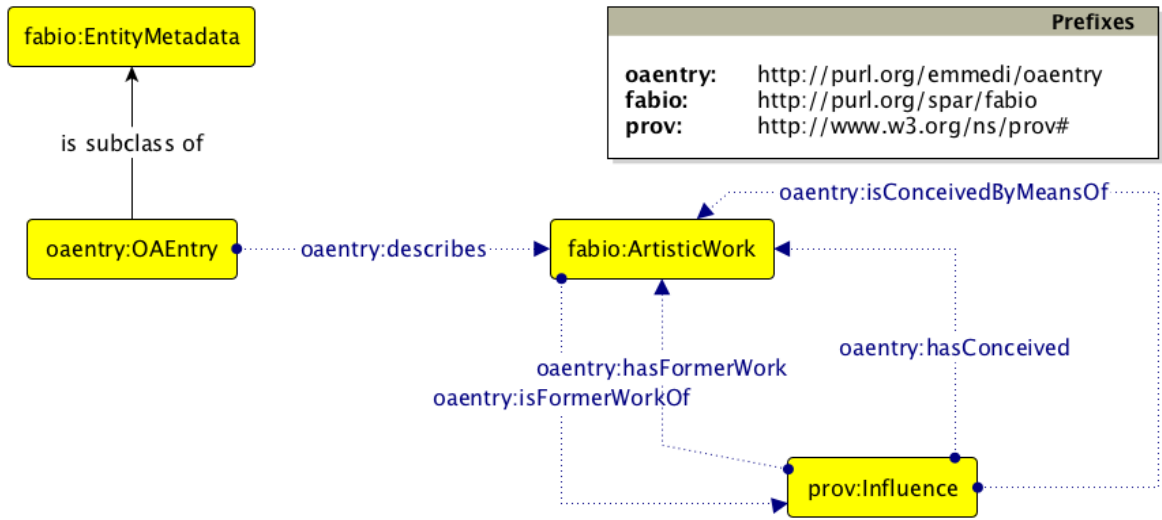

*Illustration 2: The Graffoo diagram of the OA Entry Ontology main classes and properties.*

The OA Entry Ontology introduces three main concepts: the work of art, the OA Entry that contains metadata about the work of art, and the influence between works of art. In particular:

- a work of art that can be described in different phases of its lifecycle (creation, restoration, location, ownership, custody etc.). Thus, we decided to represent a work of art as an FRBR Work when describing its essence, as an FRBR Manifestation when providing information about its physical features that may vary over time, and as an FRBR Item when dealing with information about legal aspects and its location. Modelling works of art in these terms means taking into account established considerations in the arts domain [1] [2];

- the OA entry is a document containing metadata about a work of art and its cataloguing: in the Zeri Photo Archive an OA entry is always created when an F Entry describes a work of art portrayed in a photograph. This means that some of the metadata of such entries is shared, e.g. when the entries have the same authors updating and describing entities. In our ontology, an OA entry is defined as an FRBR Work when describing its creation and an FRBR Expression when addressing issues related to its content;

- the influence between two works of art concerns the way one affects (the conception of) the other. Such influence has an effect mainly between conceptual entities (i.e., considering the FRBR Work level) rather than their concrete objects realization (i.e., the FRBR Expression level). For instance a sinopia or a preliminary drawing of a work of art is considered a proper work of art *per se*, even if it is strictly related to the final work created by an artist.

As before, in the OA Entry Ontology we reuse the models introduced for the F Entry Ontology – i.e. FaBiO, PRO, CiTO, HiCO, FRBR and PROV-O – so as to provide a precise description of specific aspects of the domain in consideration.

An OA Entry is a document containing metadata about a work of art and its cataloguing. Therefore, the OA Entry Ontology defines it in terms of an FRBR Work (i.e. an instance of the class *oaentry:OAEntry*, subclass of *fabio:EntityMetadata*). The work of art (class *fabio:ArtisticWork*) described by the entry is considered as an FRBR Work as well.

The object property *oaentry:describes* allows one to link an OA Entry to its the work of art it describes. Its range involves all the four FRBR levels, since, as introduced above, several metadata in the Zeri Photo Archive actually refer to particular aspects related to the material used for embodying the work of art (i.e., the FRBR Manifestation level) as well as its current physical location (i.e., the FRBR Item level).

## Extending PROV-O and HiCO for describing the influence between works

Original-to-derivative relations between two works of art can be represented by means of the class *prov:Influence*. In order to enable the description of different kinds of influence that can exist between works of art, the OA Entry Ontology extends PROV-O by adding appropriate subclasses to *prov:Influence* (*oaentry:Cartoon*, *oaentry:Copy*, *oaentry:Derivation*, etc.) – derived by analysing the field "ROFF" (i.e., the *status of work*) in all the metadata documents (i.e., OA entries) available in the Zeri Photo Catalogue and considering terms of the controlled vocabulary provided by the Scheda OA content standard.

The object property *oaentry:hasFormerWork* (i.e. a sub-property of *prov:entity*) allows one to link an individual of any of the influence classes to the original work of art. The object property *oaentry:hasConceived* enables one to link an individual of any of the influence classes to the derivative work of art in consideration.

By means of the aforementioned ontological entities, it is possible to properly model a scenario described in the sentence "the anonymous drawing of Sistine Chapel is conceived as a drawing of Michelangelo Buonarroti's frescos in Sistine Chapel" as shown in the following excerpt in Turtle syntax (the complete example is available at http://dx.doi.org/10.6084/m9.figshare.3175048):

```
:anonymous-drawing-sistine-chapel-work
    oaentry:isConveivedByMeansOf :michelangelo-fresco-sistine-chapel-
drawing .
:michelangelo-fresco-sistine-chapel-drawing a oaentry:Drawing ;
    oaentry:hasFormerWork :michelangelo-fresco-sistine-chapel-work ;
    oaentry:hasConceived :anonymous-drawing-sistine-chapel-work .
```

It is worth noticing that such influence assertion could also be defined as a questionable piece of information. In this case, an instance of the class *hico:InterpretationAct* can be created to specify explicitly that the claimed influence was actually derived from a cataloguer's subjective choice. To this end, the OA Entry Ontology extends HiCO by providing terms for defining types of interpretation strictly related to the arts domain – e.g. the attribution of an influence between works may be represented as the individual *oaentry:influence-between-works-attribution*. This can be defined in RDF as follows:

```
:michelangelo-fresco-sistine-chapel-drawing a oaentry:Drawing ;
    prov:wasGeneratedBy :michelangelo-fresco-sistine-chapel-drawing-
attribution .
:michelangelo-fresco-sistine-chapel-drawing-attribution a
hico:InterpretationAct ;
    hico:hasInterpretationType oaentry:influence-between-works-attribution
;
    hico:hasInterpretationType :zeri-preferred-attribution ;
    hico:hasInterpretationCriterion :cataloguer-choice ;
    hico:isExtractedFrom :oaentry-15429-expression ;
    prov:wasAssociatedWith :crr-mm .
oaentry:influence-between-works-attribution a hico:InterpretationType .
:zeri-preferred-attribution a hico:InterpretationType .
:cataloguer-choice a hico:InterpretationCriterion .
:oaentry-15429-expression a fabio:Expression .
:crr-mm a foaf:Agent .
```

## Extending PRO for providing a controlled vocabulary of roles in the Arts domain

In addition to information about a work of art, the OA Entry generally requires the cataloguer to also provide several pieces of information about the artist responsible for the creation of the work, including the actual role he/she had in the creative process. Moreover, when cataloguing the work of art, several responsibilities are attributed to cataloguers as well. We therefore added appropriate individuals to the class *pro:Role* in the OA Entry Ontology so as to describe additional roles proper to the arts and cataloguing domains. In particular, we have created two new subclasses of *pro:Role*, i.e. *oaentry:ArtisticRole* and *oaentry:CataloguingRole*, each including specific individuals:

- those defined as instances of *oaentry:ArtisticRole* have been recognized by means of the open vocabulary adopted by the Zeri Foundation for describing the roles of artists and the controlled vocabulary provided by Scheda OA, such as *oaentry:antiquarian*, *oaentry:architect*, *oaentry:art-dealer*, etc.;

- those defined as instances of *oaentry:CataloguingRole* have been recognized as the main roles involved in the cataloguing process: *oaentry:cataloguer*, *oaentry:cataloguing-institution*, *oaentry:cataloguing-supervisor*, and *oaentry:competent-institution*.

# MAPPING F/OA ENTRY TO RDF

Each Scheda F/OA content standard is organized in sections. Each section contains data about a work (a photograph, a work of art) that can be described in terms of one of the four FRBR entities (Work, Expression, Manifestation, Item).

In this section we briefly introduce the mapping we propose between the fields in the Scheda OA and Scheda F content standards to RDF according to the ontologies introduced in the previous section, and by enriching such description with terms coming from the CIDOC-CRM specification. It is worth mentioning that a complete mapping is not in the scope of this paper. In this work we focus particularly on the mapping between all the fields in Scheda F/OA that are actually used in the Zeri Photo Archive Catalogue – which includes about 118 fields out of more than 300 provided by Scheda F, and about 97 fields out of 280 provided by Scheda OA. Thus, the mappings defined includes all the mandatory elements prescribed by the standard creators (called *inventory level of description*), and a selection of other significant fields.

Operatively, one author (Marilena Daquino) performed a first round of mapping by looking at all the aforementioned fields in the Scheda F and Scheda OA content standards one by one. For each field she provided a first mapping to RDF accompanied by a meaningful example of use. After this initial process, another author (Silvio Peroni) rechecked the whole mapping document produced in order to correct mistakes, to suggest modifications, and to harmonize it where ambiguous mappings were introduced. After another iteration by the first author, so as to address all the suggestions introduced by the second, the resulting mapping document was analysed by the other three authors – i.e. a member of the Zeri Foundation responsible for Zeri's catalog (Francesca Mambelli), a digital humanist (Francesca Tomasi), and a computer scientist with a strong background in markup languages (Fabio Vitali) – so as to gather additional feedback. A new version of the mapping document was then released by the first author.

The mapping process resulted in the creation of two distinct documents, i.e. *F Entry to RDF* (https://dx.doi.org/10.6084/m9.figshare.3175273) and *OA Entry to RDF* (https://dx.doi.org/10.6084/m9.figshare.3175057), accompanied by exemplar data that represent an F Entry (http://dx.doi.org/10.6084/m9.figshare.3175252) and an OA Entry

(https://dx.doi.org/10.6084/m9.figshare.3175048) in RDF, which were created according to such mappings. All the mapping documents contain tables structured as shown in Illustration 3. Such tables reproduce the structure of the related content standard they refer to, and are organized in three columns. The first and the second column contain the name of the field in Scheda F/OA and a brief description. The third column details the mapping with RDF terms and accompanies it with examples of usage.

| OG — OBJECT AND SUBJECT DESCRIPTION * | | |
|---|---|---|
| OGT — OBJECT * | | |
| OGTD * | DEFINITION<br><br>A term identifying the main type of a described work of art. It may belong to a local open thesaurus and/or to an established one, e.g. the *Art and architecture Thesaurus*. | CRM:E55_TYPE (CLASS)<br><br>According to the *Cataloging Cultural Objects (CCO) project* of the *Visual Resources Association Foundation* (http://cco.vrafoundation.org/), which suggests to consider only the FRBR Work level when describing type of works of art, terms belonging to the open vocabulary identified in this field are considered specializations of a work of art at the FRBR Work level of description, i.e. individuals of the classes `fabio:ArtisticWork` and `crm:E28_Conceptual_Object`.<br>By means of the object property `crm:P2_has_type` a work of art may be associated to an individual defining the type of work.<br><br>EXEMPLAR USAGE:<br>`:oa-47172`<br>  `a crm:E28_Conceptual_Object , fabio:ArtisticWork ;`<br>  `crm:P2_has_type :polyptych .`<br><br>Terms of the `crm:E55_Type` hierarchy shall be aligned to an established controlled vocabulary or thesaurus, e.g. the AAT Getty Thesaurus (http://www.getty.edu/research/tools/vocabularies/aat/).<br><br>EXEMPLAR USAGE:<br>`:polyptych a crm:E55_Type ;`<br>  `rdfs:seeAlso`<br>    `<http://vocab.getty.edu/aat/300178235> .` |
| OGTT | OBJECT TYPE<br><br>A term specializing the main type of the described work of art, excluding its functional and morphological features. | CRM:E55_TYPE (CLASS)<br><br>Here, as in the OGTD field, an heterogeneous open or controlled vocabulary is used to further describe formal features of a work of art: the value of this field shall be a complementary definition of the previous one or may be considered as another specification, through the use of the property `crm:P2_has_type`.<br>E.g. OGTD: 'fountain' ; OGTT: 'basin' .<br><br>EXEMPLAR USAGE:<br>`:oa-75147`<br>  `crm:P2_has_type :basin ;`<br>  `crm:P2_has_type :fountain .`<br>`:fountain a crm:E55_Type .`<br>`:basin a crm:E55_Type .`<br><br>OR<br>`:oa-75147`<br>  `crm:P2_has_type :basin-fountain .`<br>`:basin-fountain a crm:E55_Type .` |

*Illustration 3: An excerpt of the mapping document "Mapping OA Entry to RDF".*

The following sections are organized on the basis of the four FRBR levels, with an introduction about the representation of top-level relations between entries and subjects described therein. The ontological entities related with the F Entry Ontology and the OA Entry Ontology (presented in THE F ENTRY ONTOLOGY AND THE OA ENTRY ONTOLOGY) are directly used without further explanation, while the use of CIDOC-CRM in the RDF excerpts is detailed.

## *The entry and its subject*

Any F/OA Entry can be defined in CIDOC-CRM as an instance of the class *E31 Document*, and it is related to its respective subjects by means of the object property *P70 documents*. An explicit relation between the F Entry describing a photograph and an OA Entry describing the work of art portrayed in that photograph can be represented by using *P67 refers to*.

Identifiers of the entries are represented by means of an instance of the class *E42 Identifier*. An entry can have several identifiers, and all the identifiers can be represented in the same way and then characterized with the property *P2 has type*, so as to specify which type of identifier is associated (levels of cataloguing, regional codes, general catalogue numbers, etc.).

We decided to use terms belonging to PRO as well as to the individuals of the class *oaentry:CataloguingRole* for describing the cataloguing process, which involves different roles at different times.

The following Turtle excerpt below provides an example of all the aforementioned aspects:

```
:fentry-72486 a fentry:FEntry , crm:E31_Document ;
      fentry:describes :photo-72486 , :oa-47172 ;
      crm:P67_refers_to :oaentry-43677 .
:fentry-72486-creation a crm:E65_Creation ;
      crm:P14_carried_out_by :cataloguer ;
      crm:P4_has_time_span :2016-
:oaentry-43677 a oaentry:OAEntry , fabio:Work , crm:E31_Document ;
      oaentry:describes :oa-47172 ;
      crm:P140i_was_attributed_by
            :oaentry-43677-catalog-level-assignment ,
            :oaentry-43677-nctr-assignment , :oaentry-43677-nctn-
assignment ;
:md-cataloguer-oaentry-43677 a pro:RoleInTime ;
      pro:relatesTo :oaentry-43677 ;
      pro:isHeldBy :md ;
      pro:withRole oaentry:cataloguer ;
      tvc:atTime :2012-11-04 .
```

## *The Work level*

Both the photograph (described in an F Entry) and the work of art (described in an OA Entry) are defined in CIDOC-CRM terms as instances of the class *E28 Conceptual Object* when describing their essence and their creation.

The class *E65 Creation* is required to define the authorship of photographs and works of art, i.e. to identify the main photographer of the photograph and the main artist or group of artists of the work of art. The actors (*E39 Actor*, or one of its subclasses, i.e., *E21 Person* and *E74 Group*) involved in this authorship attribution are specified by using the object property *P14 carried out by*. The object property *P4 has time span* is used for specifying the duration of the creation event.

Moreover, the shot can be associated to a place (the class *E53 Place*) by using the object property *P8 took place at*, and can be associated to a specific occasion, i.e. an instance of the class *E4 Period* (further specified in *E5 Event*) by means of the object property *P10 falls within*.

Both the creators of the two aforementioned works can be associated to a cultural context (for instance a school of painters or a workshop) by using the property *P107i is current of former member of*, which relates them to an individual of the class *E74 Group*.

According to Bountouri and Gergatsoulis [3] [4], the archival description of the photograph, i.e. the hierarchical organization of the containers that include the catalogued object, can be described as nested relations by using the object property *P106 is composed of* for relating conceptual entities.

One or more titles can be attributed to the entities by means of the object property *P102 has title* associated to an instance of the class *E35 Title,* which can be further specialized by using *P2 has type* to define if the title is an attributed, traditional or an alternate one.

When a bibliography or other sources are provided to support the cataloguing, e.g. letters, audio-recorded works, catalogues, entries etc., a generic relation can be represented with *P70i is documented in*, by linking to an individual of the broader class *E31 Document*.

A direct relation between the photograph and the depicted work of art can be established relating the photograph (the FRBR Work) to the concrete object of art (the FRBR Item).

The following Turtle excerpt introduces an example of part of the aforementioned aspects:

```
:photo-72486 a fentry:Photograph , crm:E28_Conceptual_Object ;
    crm:P94i_was_created_by :photo-72486-creation ;
    crm:P106i_forms_part_of :folder-leonardo ;
    crm:P102_has_title :jesus-baptism-verrocchio ;
    frbr:subject :oa-47172-item ;
    crm:P70i_is_documented_in :document-f2336 .
:photo-72486-Creation a crm:E65_Creation ;
    crm:P94_has_created :photo-72486 , :photo-72486-expression ;
    crm:P7_took_place_at :florence ;
    crm:P10_falls_within :exhibition-of-paintings ;
    crm:P4_has_time_span :1926-1932 ;
    crm:P14_carried_out_by :brogi-studio .
:folder-leonardo a fabio:Work , crm:E90_Symbolic_Object ;
    crm:P106i_forms_part_of :subseries-leonardo .
:subseries-leonardo a fabio:Work , crm:E90_Symbolic_Object ;
    crm:P106i_forms_part_of :series-leonardo .
:series-leonardo a fabio:Work , crm:E90_Symbolic_Object ;
    crm:P106i_forms_part_of :zeri-photo-archive .
:zeri-photo-archive a fabio:WorkCollection .
:oa-47172 a fabio:ArtisticWork , crm:E28_Conceptual_Object ;
    crm:P94i_was_created_by :oa-47172-creation ;
    fabio:hasPortrayal :oa-47172-item .
```

## *The Expression level*

While in the arts domain we are not interested in representing the contents of a work of art separately from their conception, in the photography domain we can describe the photograph with regards to its content, i.e. its FRBR Expression, which is realized at the same time as the creation of the work. None of this information is precisely covered by the CIDOC-CRM, then terms from FaBiO and F Entry Ontology are used.

The following Turtle excerpt introduces an example of the aforementioned aspects:

```
:photo-72486-creation a crm:E65_Creation ;
    crm:P94_has_created :photo-72486 , :photo-72486-expression .
:photo-72486-expression a fentry:Shot ;
    crm:P94i_was_created_by :photo-72486-creation ;
    frbr:realizationOf :photo-72486 .
```

## *The Manifestation level*

The description about the format used for embodying an object applies to the Manifestation level.

On the one hand, in photographs different formats identify several manifestations, e.g. digital images, slides, negatives, and positives. Each manifestation belongs also to the class *E22 Man-Made Object* and represents a specific form that the work may have, which can be further characterised by means of the object property *P2 has type*.

On the other hand, in works of art the FRBR Manifestation level should be applied any time a relevant change affected the object, e.g. any time a restoration intervention was recorded.

Both the photograph and the work of art may be described in terms of the material they are made of, i.e. an instance of the class *E57 Material*. The class *E16 Measurement* is used to define the various dimensions (*E54 Dimension*) related with such manifestations (e.g. weight and height). Other specific features characterizing the manifestation, e.g. colour, are linked with the object property *P56 bears feature*.

The following Turtle excerpt introduces an example of all the aforementioned aspects:

```
:photo-72486-positive
    a crm:E22_Man-Made_Object , fabio:AnalogManifestation ;
    crm:P45_consists_of :gelatin-silver ;
    crm:P56_bears_feature :black-and-white ;
    crm:P39i_was_measured_by :photo-72486-positive-measurement .
:photo-72486-positive-measurement a crm:E16_Measurement ;
    crm:P40_observed_dimension :height-194mm ;
    crm:P40_observed_dimension :width-250mm .
```

### *The Item level*

The photograph is considered as a concrete object that can be defined as an individual of the class *E22 Man-Made Object* (also inferred as *E84 Information Carrier*) linked to the portrayed work of art by means of the object property *P62 depicts*.

Specific features regarding the object reported by cataloguers during an assessment (represented with the class *E14 Condition Assessment*) are defined as instances of the class *E3 Condition State*. It is worth mentioning that any instance of *E3 Condition State* can be further specialized by using the *P2 has type* property, e.g. for defining the status of the object recorded during the assessment.

Several situations may be related to the production of the concrete object representing the photograph or the work of art, where several actors with a specific role may be involved. Therefore terms belonging to PRO Ontology are here preferred.

Different sorts of location (the class *E53 Place*) can be attributed to a concrete object by using the object property *P55 has current location*. When filling in the section regarding the geographical and administrative location of an object (either a photograph, a work of art or a work related to either), a cataloguer can refer to a place, i.e. class *E53 Place*, and/or to a current keeper (*P50 has current keeper*) which resides in such a place. In the latter case, the current keeper (*E39 Actor*) is further related to a place by using the object property *P74 has current or former residence*, and to an address by using the property *P76 has contact point*. The place in which the keeper resides is defined by means of a chain of nested places (e.g. village, town, district, region, country), each other related through the object property *P89 falls within*.

Such complete description of the current location of an object, i.e. in terms of a place and an agent keeping it, enables the description of its transfers of custody (*E10 Transfer of Custody*). If a particular keeper is not identified, the change of location can be simply defined in terms of *E9 Move*, linked the location the object has been moved (*P26 moved*) from another one (*P27 moved from*).

The ownership of the concrete object can be defined by using the property *P52 has current owner*. Each owner (*E39 Actor*) could have acquired the work (property *P22i acquired title through*) as the consequence of an acquisition event *E8 Acquisition*, that can be further specialized by using *P2 has type* to describe ownership specifications.

The last considered section regarding the concrete object addresses its participation in exhibitions (*E5 Event*). Such events can, thus, be linked to the object included in the event (*P12 occurred in the presence of*) to the location (*P7 took place at*) and date (*P4 has time span*) of the event, which can have a formal appellation (*E41 Appellation*) specified through the property *P1 is identified by*.

The following Turtle excerpt introduces only a partial example of the aforementioned aspects regarding a photograph, since a work of art ca be described similarly:

```
:photo-72486-positive-item a fabio:AnalogItem , crm:E22_Man-Made_Object ;
    crm:P62_depicts :oa-47172-item ;
    crm:P57_has_number_of_parts "1" ;
    crm:P34i_was_assessed_by :photo-72486-positive-item-condition ;
    crm:P55_has_current_location :large-formats-room ;
    crm:P140i_was_attributed_by :photo-72486-invn-assignment ;
    crm:P52_has_current_owner :university-of-bologna ;
    crm:P12i_was_present_at :exhibition-london-1987 ;
    crm:P30i_custody_transferred_through :photo-72486-item-provenance-1 .
:photo-72486-positive-item-condition a crm:E14_Condition_Assessment ;
    crm:P35_has_identified :photo-72486-positive-item-condition-state .
:photo-72486-positive-item-condition-state a crm:E3_Condition_State ;
    crm:P2_has_type :discrete ;
    crm:P3_has_note "silver mirror" .
:large-formats-room a crm:E53_Place ;
    crm:P89_falls_within :ex-convent-santa-cristina .
:photo-72486-item-provenance-1 a crm:E10_Transfer_of_Custody ;
    crm:P28_custody_surrendered_by :villa-i-tatti ;
    crm:P29_custody_received_by :zeri-foundation ;
    crm:P30_transferred_custody_of :photo-72486-positive-item ;
    crm:P4_has_time_span :1989 .
```

## THE ZERI PHOTO ARCHIVE RDF DATASET

The specific heterogeneous nature of the domains considered in the *Zeri and LODE* project (and in the PHAROS project) and the lack of similar Linked Open Data datasets available on the Web were important issues to address in order to provide a comprehensive description of these different kinds of information by means of Semantic Web technologies. Thus, we believe that publishing a dataset about these kinds of objects could be considered a pioneering attempt *per se*, and would represent, in principle, an established practice for creating a network of LOD with other Italian cultural institutions. In addition, it can be considered an available source of precious information for linking existing international datasets in the field of the Arts.

The Zeri Photo Archive RDF dataset we have realized (available online at https://w3id.org/zericatalog/), which contains data compliant with the ontologies and the mappings described in the previous sections, is the result of an analysis of a dataset originally created for testing purposes in the context of the PHAROS project. More specifically, we gathered about 31.000 F Entries and 19.000 OA Entries from the Zeri Photo Archive catalogue, stored as XML documents (compliant with no particular schema) that contained the data prescribed by the Scheda F and Scheda OA content standards. XML contents were organized in *paragraphs*, which correspond to Scheda F/OA descriptive sections. The data gathered were chosen according to a thematic organisation. The first subset of the Zeri catalogue considered was based on a collection of OA Entries describing works of the fifteenth and sixteenth centuries, and included all the F Entries describing photographs portraying the aforementioned works.

The resulting RDF dataset was created by means of an XSL transformation of the provided XML documents into RDF statements – where one RDF/XML file has been created for each of the considered entries. RDF statements mainly refer to photographs and works of art, and also include relevant information describing about 4,500 bibliographic entities, 6,000 artists, 2,000 photographers and 2,000 catalogues. This additional information was provided by the

Zeri Foundation by means of other XML documents representing local authority files. These XML files reproduce the fields required by the ICCD guidelines for creating people authority files (http://www.iccd.beniculturali.it/index.php?it/473/standard-catalografici/Standard/55) and bibliographic references authority files (http://www.iccd.beniculturali.it/index.php?it/473/standard-catalografici/Standard/58). The aforementioned authority files were converted into RDF (by creating one RDF/XML file for each record in an authority file) by means of an XSL transformation.

All the RDF resources were labelled in Italian and, where possible, in English, in order to facilitate their understanding for a larger audience. In addition, the IRIs of these resources were created in English in order to ensure their easy reuse in other non-Italian datasets. The document describing the IRI design choices is available at the homepage of the project, (https://w3id.org/zericatalog, section *Data*).

Several open vocabularies adopted by cataloguers in original data were aligned to closed controlled vocabularies, and all the RDF resources were further enhanced with direct links to external authority files and datasets, in particular:

- the Getty Art & Architecture Thesaurus (http://www.getty.edu/research/tools/vocabularies/aat/) was considered for aligning terms referring to object types, materials and techniques;
- VIAF (http://viaf.org), Getty Union List of Artist Names (http://www.getty.edu/research/tools/vocabularies/ulan) , Wikidata (https://www.wikidata.org), and Dbpedia (http://dbpedia.org) were chosen to identify artists and photographers (the latter still in progress);
- geoNames (http://www.geonames.org) and, again, Wikidata and Dbpedia were considered for identifying places.

Reconciliation of entities to established authority files like VIAF and geoNames was mainly obtained through a semi-automatic process, by using ad hoc plugins implemented in OpenRefine (http://openrefine.org/) . Moreover, a PERL script (http://search.cpan.org/dist/App-wdq/lib/App/wdq.pm) was useful for accessing the Wikidata Query Service, which also enabled us to directly link entities to Wikipedia pages, Dbpedia and Getty ULAN entities.

All the RDF statements were stored in an Apache Jena Fuseki2 triplestore (https://jena.apache.org/documentation/fuseki2). We chose Fuseki2 because it is easy for even non-expert users to deploy and manage, but we will be probably have to change it by the end of the PHAROS project, due to its limits in managing huge amounts of data. All the data stored in the triplestore are distributed under the license CC-BY-NC, Creative Commons Attribution-NonCommercial 4.0 International (http://creativecommons.org/licenses/by-nc/4.0/).

As of September 11, 2016, the dataset includes about 11,400,000 RDF statements relating 1,600,000 unique typed entities. Among these, about 3,000 have been linked with external resources already available in the Linked Open Data. In particular, we created links to 2,200 different VIAF resources, 1,200 to Getty ULAN resources, 1,500 different GeoNames resources, and 2,260 different Dbpedia and Wikidata resources. We estimate that we will have 100 million RDF statements in the dataset once the full Zeri catalogue has been processed.

All the RDF data can be queried in SPARQL by making appropriate REST requests to the related SPARQL endpoint made available by the triplestore at https://w3id.org/zericatalog/sparql. We have also prepared a web interface (available at http://data.fondazionezeri.unibo.it/query/) for allowing users to query the triplestore directly on the Web by means of SPARQL queries. Finally, we have used LODView (http://lodview.it/)

for to allow direct browsing of all the RDF data included in the triplestore – for instance, Illustration 4 shows the HTML visualisation of the resource https://w3id.org/zericatalog/photo/59972, i.e. the photograph "Alinari, Fratelli , Perugia, Collegio del Cambio: Cappella di S. Giovanni Battista, volta - S. Giovanni e S. Luca (Giannicola di Paolo)".

All F Entries and OA Entries defined in the dataset include links to the current photograph and arts catalogue of the Zeri Foundation (available at http://catalogo.fondazionezeri.unibo.it), enabling users to go from the LOD-based representation of the catalogue to the traditional Web pages which present the same entities. The inverse link from HTML pages to the related RDF resources has been implemented and is currently in a test phase (and not yet officially released).

*Illustration 4: An example of a resource of the Zeri Photo Archive visualized in LODView.*

# CONCLUSIONS

In this article we have introduced the main outcomes to date of the *Zeri and LODE* project: the F Entry Ontology, the OA Entry Ontology, the mapping documents between two Italian content standards (Scheda F and Scheda OA) and CIDOC-CRM accompanied by the aforementioned ontologies, and the creation of a real LOD dataset listing about 50,000 entries from the Zeri Photo Archive. Our project is set within the PHAROS project, which brings together an international consortium of several institutions to create a common platform for research on the images and metadata of Western and non-Western works.

As required by the PHAROS Consortium, the *Zeri and LODE* project have already produced a first stable outcome that allows the description of a large set of metadata about photographs and works of art. In the next phase, we will drive the adoption of these ontologies by other institutions (a dialogue with ICCD has already been set up so as to work towards a strategy for the adoption of our models at a national level) and we will extend the current LOD dataset to include all of the data expressed in the other F/OA entries in Zeri's catalogue, so as to make available openly and in machine-readable form the full richness of Federico Zeri's knowledge base. Of course, this extension will require a new hardware architecture – e.g. a powerful new server, the use of a more performant triplestore such as Blazegraph (https://www.blazegraph.com/), as well as additional disk space and RAM for storing and managing the new data – which is something we will deal with in the near future.

At the same time, the ontologies and the corresponding mapping documents we have developed will be revised and extended in order to complete the analysis of the F/OA contents standards, also considering relevant topics introduced in the earliest version of the Scheda F (version 4.00 at http://www.iccd.beniculturali.it/index.php?it/473/standard-catalografici/Standard/62). Therefore a complete mapping will be provided to allow the most comprehensive description of the Photography and Arts domain as addressed in ICCD content standards.

It is worth mentioning that, even if the mapping we provided was based on national content standards and meant comparing and merging two slightly different perspectives (i.e. the official cataloguing standard requirements and its implementation in a real use case where some customizations and a few new fields were defined by the Zeri Foundation itself), we believe it could be considered an exhaustive resource for any international cultural heritage institution wanting to publish its data as LOD. The large number of models used to map such standards into RDF was useful for defining a coherent and comprehensive descriptive scenario, based on state-of-the-art ontologies, for the cultural heritage domain.

In addition, it is possible that future revisions of the ontologies and of the mapping documents will take into account additional models that were not considered at this stage, such as FRBRoo. While FRBRoo matches well with CIDOC-CRM by design, its adoption in our project was not part of the requirements given by the PHAROS Consortium. In addition, at the current stage of our project and because of the domain we are describing, FRBRoo would not add any meaningful additional detail to the current data, and thus we decided to postpone its use into future phases of the project – when at least a formal alignment to it would be provided. In addition to FRBRoo, we are planning to include proper models for archival description, e.g. the Reload Ontologies (http://labs.regesta.com/progettoReload/), extensively adopted in Italian archives, in order to enable an easier integration of the Zeri catalog to LOD datasets made available by other archival institutions.

Finally, once other partners of the PHAROS Consortium publish their data according to the defined models, the second phase of the project will allow us to integrate all of the data in a common environment, together with a searchable repository for images (at the moment, a beta version of this application to search images is available at http://en.images.pharosartresearch.org/). The aim is to combine current search capabilities, which enable the user to find similar images across the data of participating institutions, with a powerful search on metadata, enabling the user to retrieve different (and contradictory) descriptions about the same artwork.

## ACKNOWLEDGEMENTS

We would like to thank Ciro Mattia Gonano, who first addressed relevant issues in mapping Scheda F to CIDOC-CRM and whose work was the basis for the current evolution of the project, and Raffaele Messuti, who helped us in managing the current RDF dataset, deploying and maintaining the server, and in reconciling data to external vocabularies and datasets. We would also like to thank the editor and the reviewers of this paper, who have provided us with valuable comments and suggestions for improving the article.